\documentclass[graybox]{svmult}

\usepackage{type1cm}        
\usepackage{makeidx}         
\usepackage{graphicx}        
\usepackage{multicol}        
\usepackage[bottom]{footmisc}

\usepackage{newtxtext}       %
\usepackage{newtxmath}       

\usepackage{cite}
\usepackage[normalem]{ulem}

\newcommand{\tb}[1]{\textcolor{black}{#1}}
\newcommand{\mb}[1]{\textcolor{black}{#1}}
\newcommand{\tbnotes}[1]{\textit{\textcolor{red}{#1}}}
\renewcommand{\tbnotes}[1]{}
\renewcommand{\sout}[1]{}

\makeindex             

\begin{document}

\title*{Modeling of obstacle avoidance by a dense crowd as a Mean-Field Game}
\author{Matteo Butano, Thibault Bonnemain, Cécile Appert-Rolland, Alexandre Nicolas, Denis Ullmo}
\authorrunning{Butano, Bonnemain, Appert-Rolland, Nicolas, Ullmo}
\institute{Matteo Butano \at Universite Paris Saclay, CNRS, LPTMS, 91405, Orsay, France \\ \email{matteo.butano@universite-paris-saclay.fr}
\and Thibault Bonnemain \at Department of Mathematics, King's College, London, United Kingdom 
\and Cécile Appert-Rolland \at Université Paris-Saclay, CNRS, IJCLab, 91405, Orsay, France
\and Alexandre Nicolas \at Institut Lumière Matière, CNRS \& Université Claude Bernard Lyon 1, 69622, Villeurbanne, France
\and Denis Ullmo \at Universite Paris Saclay, CNRS, LPTMS, 91405, Orsay, France \\ \email{denis.ullmo@universite-paris-saclay.fr}
}
%
%
\titlerunning{Modeling of obstacle avoidance by a dense crowd as a Mean-Field Game}

\maketitle

\abstract*{In this paper we use a minimal model based on Mean-Field Games (a mathematical framework apt to describe situations where a large number of agents compete strategically) to simulate the scenario where a static dense human crowd is crossed by a cylindrical intruder. After a brief explanation of the mathematics behind it, we compare our model directly against the empirical data collected during a controlled experiment replicating the aforementioned situation. We then summarize the features that make the model adhere so well to the experiment and clarify  the anticipation time in this framework. }

\abstract{In this paper we use a minimal model based on Mean-Field Games (a mathematical framework apt to describe situations where a large number of agents compete strategically) to simulate the scenario where a static dense human crowd is crossed by a cylindrical intruder. After a brief explanation of the mathematics behind it, we compare our model directly against the empirical data collected during a controlled experiment replicating the aforementioned situation. We then summarize the features that make the model adhere so well to the experiment and clarify  the anticipation time in this framework.  }

\section{Introduction}
\label{sec:intro}
\tbnotes{It will depend on the space we have left but if we can I would add a few references...}
It has been established that the motion of individuals in crowded environments can be regarded at different scales. In fact, following the classification of \cite{hoogendoorn2004pedestrian}, it is possible to distinguish  between the \textit{strategic} level, describing the goal of the journey, where a certain individual is heading to; the \textit{tactical} level, or which route will be chosen to reach the final destination; and finally the \textit{operational} level, the actual trajectory along the chosen route. In the present Paper, we focus on the operational aspects of  pedestrians' motion in dense crowds, and in particular we study how individuals react to the presence of a moving cylindrical intruder. 

\begin{figure}
    \sidecaption
    \includegraphics[width=6.5cm]{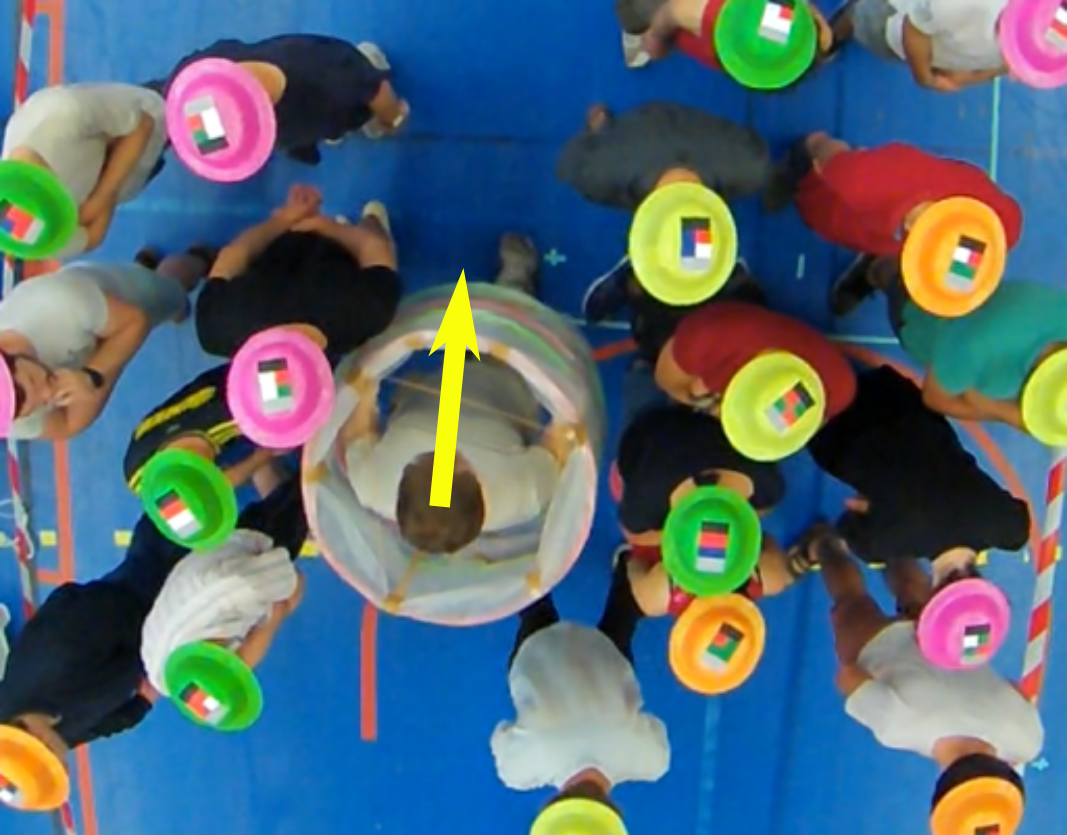}
    \caption{Photograph of the experimental configuration, with a cylindrical intruder passing through a crowd with a density $2.5$ $\text{ped}/m^2$. Pedestrians' motion was tracked with the help of colored hats and then transformed into the numerical data shown in the next figure\tb{, after averaging over sufficiently many realisations}. }
    \label{fig:exp_config}
\end{figure}

\begin{figure}
    \centering
    \includegraphics[width= 0.8\linewidth]{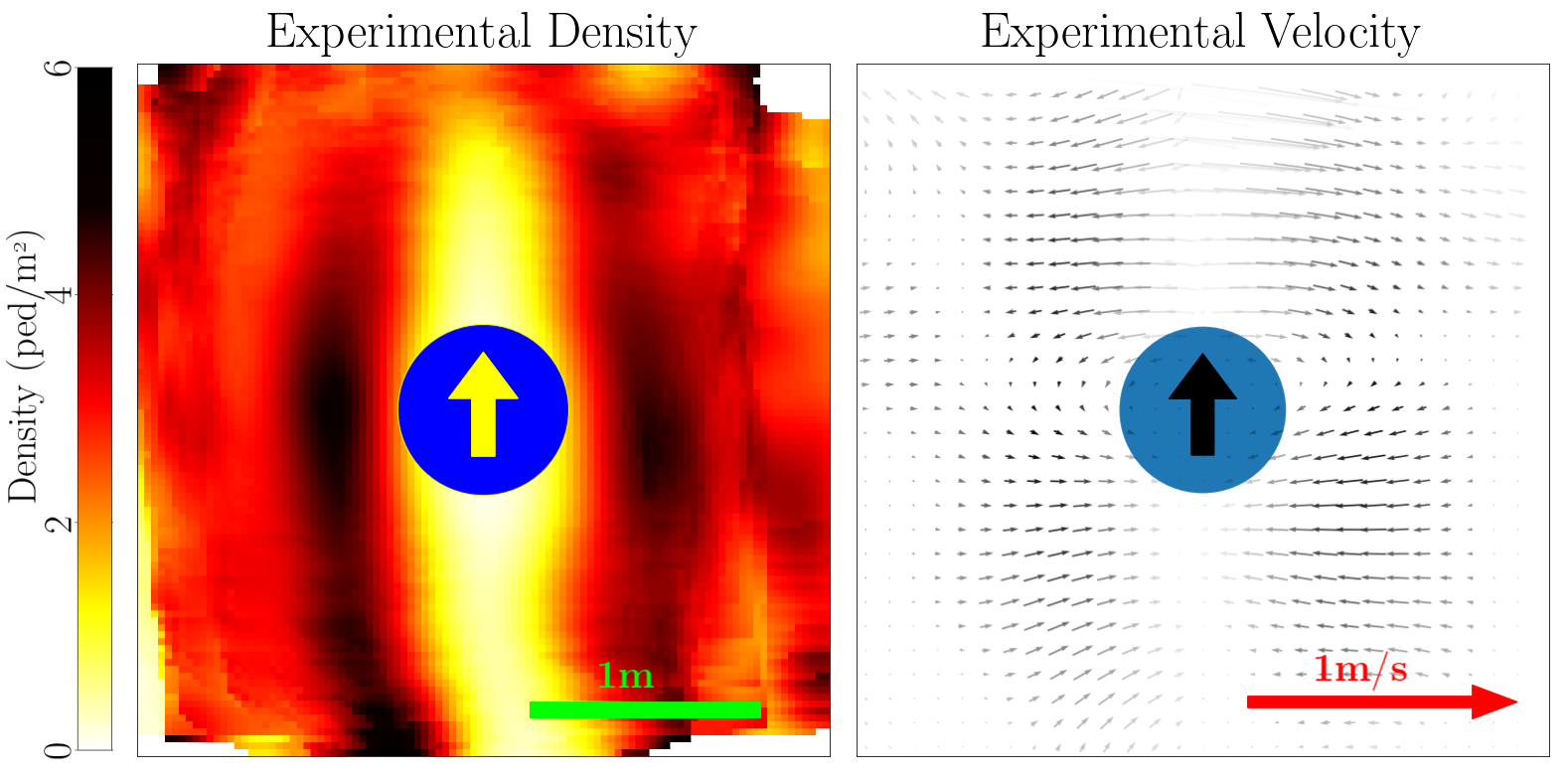}
    \caption{\tb{Averaged d}ensity and velocity fields of \tb{pedestrians in} the experiment reported in \cite{Nicolas_Kuperman_mech_response} for a \tb{mean} density of 2.5 ped$/m^2$. Pedestrians step aside, accepting to move to a higher density region \tbnotes{although it is true that this minimises the *overall* discomfort, presenting it like at this point of the paper (before explaining MFG) is confusing and (I feel) not very pedagogical... although there is probably a way to say something similar but with a different wording and a longer explanation, it will depend on the space we have left.} \tb{before} return\tb{ing} to a calmer area once the \tb{intruder} has passed.}
    \label{fig:exp_results}
\end{figure}

The choice of such a scenario is motivated by a recent series of experiments where  people were gathered in a controlled environment to form crowds with different average densities, that were crossed by a cylindrical intruder \cite{Nicolas_Kuperman_mech_response,TGF2018}. Figure \ref{fig:exp_config} shows the experimental configuration, where we also notice that participants were wearing colored hats to record their  displacement.
As we can see from the empirical results plotted in Figure \ref{fig:exp_results}       
pedestrians step aside well in advance of the cylinder's arrival, inducing a higher density at its sides, but a depletion in front of and behind it. Individuals accept to temporarily move towards a higher density area,  \tb{to let the intruder through}, and then readily head \tb{back} towards a less crowded. It is remarkable that this behavior is observed for densities ranging from 2.5 up to 6 ped$/m^2$.

These data motivated the \tb{inquiry into} a model that would be able to replicate the main qualitative \tb{features observed experimentally.} \tb{In the present Paper we argue that a minimal model based on Mean-Field Games is the right choice for this endeavour.}

\section{The Mean-Field Games model}
\label{sec:mfg}
Mean-Field Games (MFG) were introduced by J.-M. Lasry and P.-L. Lions \cite{Lasry_Lions_horizon_fini,Lasry_Lions_stationnaire}, and by M. Huang, R. P. Malhamé and P. E. Caines  \cite{Huang_Malhamé_large_pop_stoch_dyn}, \tb{to deal with optimisation problems comprising a large number of interacting agents}. Much has been said about the mathematics of MFG \cite{Cardaliaguet_notes_mfg,Gomes_Saude_mfg_survey,Bensoussan_Frehse_master_eq_mft}  \tb{but}, although applications to pedestrian dynamics  \tb{have been considered} \cite{Lachapelle_Wolfram_mfg_cong,djehiche2017mean, nasser2019crowd}, to the best of our knowledge  direct comparisons between MFG and \tb{(}real world\tb{) experimental} data \tb{regarding the local navigation of crowds} have \tb{yet to be performed}.

\subsection{The mathematical model}
\label{subsec:mfg_model}
\tb{We refer the reader to \cite{Cardaliaguet_Delarue_master_eq_convergence_prob} for a} general and mathematically rigorous discussion of the foundations of MFG, \tb{and to \cite{Ullmo_Swiecicki_quadratic_mfg} for} a physicist-friendly \tb{introduction}. For our purposes it suffices to know that MFG are  optimally driven diffusive processes \tb{involving many interacting} agents. More explicitly, \tb{let us} consider a \tb{time-dependent} \emph{differential game}\footnote{A differential game is a game involving continuous state variables.} played by a large number of agents $N\gg1$. At \tb{any} time $ t $, \tb{one} can associate to \tb{every} agent \tb{$i$ a} \emph{state variable,} $ \vec{X}_i(t)\in \mathbb{R}^2$, \tb{relevant to the game:} \tb{in this case a position}. \tb{Over the duration $T$ of the game (e.g. the time the intruder needs to go through the whole crowd), we assume agents are able to control their desired velocity}\footnote{\tb{We keep here the canonical notation $\vec{a}_i(t)$ associated with control parameters in game theory.}}  \tb{ $ \vec{a}_i(t) \in \mathbb{R}^2$, such that we can represent their individual motion via \emph{Langevin dynamics}} 
\begin{equation}
	\label{eqn:langevin}
	\dot{\vec{X}}_i = \vec{a}_i(t)+\sigma_i\vec{\xi}_i(t),
\end{equation}
 where $\vec{\xi}_i(t)$ is a $2$-dimensional vector of uncorrelated Gaussian white noises \tb{accounting for phenomena agents cannot predict or control}. \tb{Agents decide on a strategy (i.e. the choice of $\vec{a}_i(t)$ for all $t\in [0,T]$)  in order to minimise their individual cost functional $c_i$. Choosing the correct form of $c_i$ is the \mb{ key aspect} of our game theoretical approach, as it informs all the physics. Although a more refined version could be considered, we argue that the following minimal model is sufficient to reproduce the experimentally observed behaviour.} 
\begin{equation}
	\label{def:cost}
	c_i[\vec{a}_i] (\vec{X},t)=\mathbb{E}\left\{\int_t^T\left[ \frac{\mu_i}{2}(\vec{a}_i(\tau))^2 - V_i(\vec{X}(\tau),\tau)\right]d \tau\right\},
\end{equation}
 where $\vec{X}(t) = (\vec{X}_1(t),\dots,\vec{X}_N(t)) $ \tb{and $\mathbb{E}$ denotes averaging over realisations of the noise}. \tb{The first term of the integrand plays the role of a kinetic energy and represents the efforts made by the agent to enact their strategy: having to move is uncomfortable and rushing is even more. The second, $V(\vec{X},\tau)$, is a potential that describes how agents interact with each other and with the environment, which we shall soon define}.

\tb{For a large number of players, the problem is essentially intractable (see \cite{hoogendoorn2003simulation} for an example of optimal control applied to small crowds), and we need to} make some \tb{key} assumptions. The first one is that all players are identical \tb{up to their initial position: $\{\forall i=1\dots N,\, \mu_i=\mu, \, V_i = V, \sigma_i=\sigma\}$. The second is that the potential depends on the agents' positions only through the empirical density $\tilde{m}(\vec{x},t) =\frac{1}{N}\sum_{i = 1}^{N}\delta(\vec{x}-\vec{X}_i(t))$, which in the presence of many agents self-averages to $m(\vec{x},t) \equiv \mathbb{E}[ \tilde{m}(\vec{x},t) ]$. This is the eponymous mean-field approximation, agents only interact through the density $m(\vec{x},t)$. Finally we assume the potential to be of the simple form
\begin{equation}
	\label{def:potential}
	V(\vec{X}(t),t) = V[m](\vec{x},t) = gm(\vec{x},t) + U_0(\vec{x},t).
\end{equation}}
\tb{The first term in the right hand side accounts for interactions between agents: choosing $g<0$ amounts to penalising agents for standing in an overcrowded area. The second term represents interactions with the environment, where $U_0(\vec{x},t) \equiv V_0 \Theta(||\vec{x}-\vec{v}t|| - R)$,  $V_0 \to -\infty$, to model the cylindrical intruder of radius $R$ and velocity $\vec{v}$. Ultimately each agent strives to optimise the same simplified cost functional}
\begin{equation}
	\label{def:mean_cost}
	c[\vec{a}](\vec{x},t) = \mathbb{E}\left\{\int_t^T\left[ \frac{\mu}{2}(\vec{a}(\tau))^2 - V[m](\vec{x},\tau)\right]d \tau\right\},
\end{equation}
\tb{given their initial position $\vec X_i(t=0)$.}
\tb{Introducing the \emph{utility function}} $u(\vec{x},t) \equiv \min_{\vec{a}}c[\vec{a}](\vec{x},t)$, the \textit{dynamic programming principle} \cite{Bellman_dynamics_programming} \tb{implies $u(\vec{x},t)$ solves the  \textit{Hamilton-Jacobi-Bellman} (HJB) equation
\begin{equation}
	\label{eqn:HJB_disc}
		 \partial_t u + \frac{\sigma^2}{2}\Delta{u} + \min_{\vec a}\left[\frac{\mu}{2}(\vec a)^2 + \vec a . \vec \nabla u\right] =  V[m] ,
\end{equation}}
\tb{where minimisation of the left hand side then yields the optimal control $\vec{a}^* = -{\vec{\nabla}u}/{\mu}$. Self-consistency of the mean-field approximation is ensured by having the density of agents -- individually following Langevin dynamics \eqref{eqn:langevin} -- solve a \emph{Kolmogorov-Fokker-Planck} (KFP) equation, but one where the drift term is the optimal control $\vec{a}^*$. Hence, solving the MFG problem amounts to solving the system of coupled PDEs
\begin{equation}
	\label{eqn:MFG_sys}
	\begin{cases}
		\partial_t u + \frac{\sigma^2}{2}\Delta{u} - \frac{1}{2\mu}(\vec{\nabla}u)^2 =  V[m] \quad &\text{[HJB]}\\
		\partial_t m = \frac{\sigma^2}{2}\Delta m  +\frac{1}{\mu}\nabla\cdot(m\nabla u) \quad &\text{[KFP]}
	\end{cases}
 .
\end{equation}}
\tb{We should stress that HJB equation has terminal condition $u(\vec{x},t=T)=0$ given by definition \eqref{def:mean_cost}, while KFP has initial condition $m(\vec{x},t = 0) = m_0$, assuming agents are initially uniformly distributed\footnote{This is the optimal configuration given only repulsive interactions, before arrival of the intruder.}. This \emph{forward-backward} structure has profound implications on the game dynamics and encodes the agents' ability to plan ahead.}

\subsection{The simulation}
\label{subsec:mfg_results}
\begin{figure}
    \centering
    \includegraphics[width=0.8\linewidth]{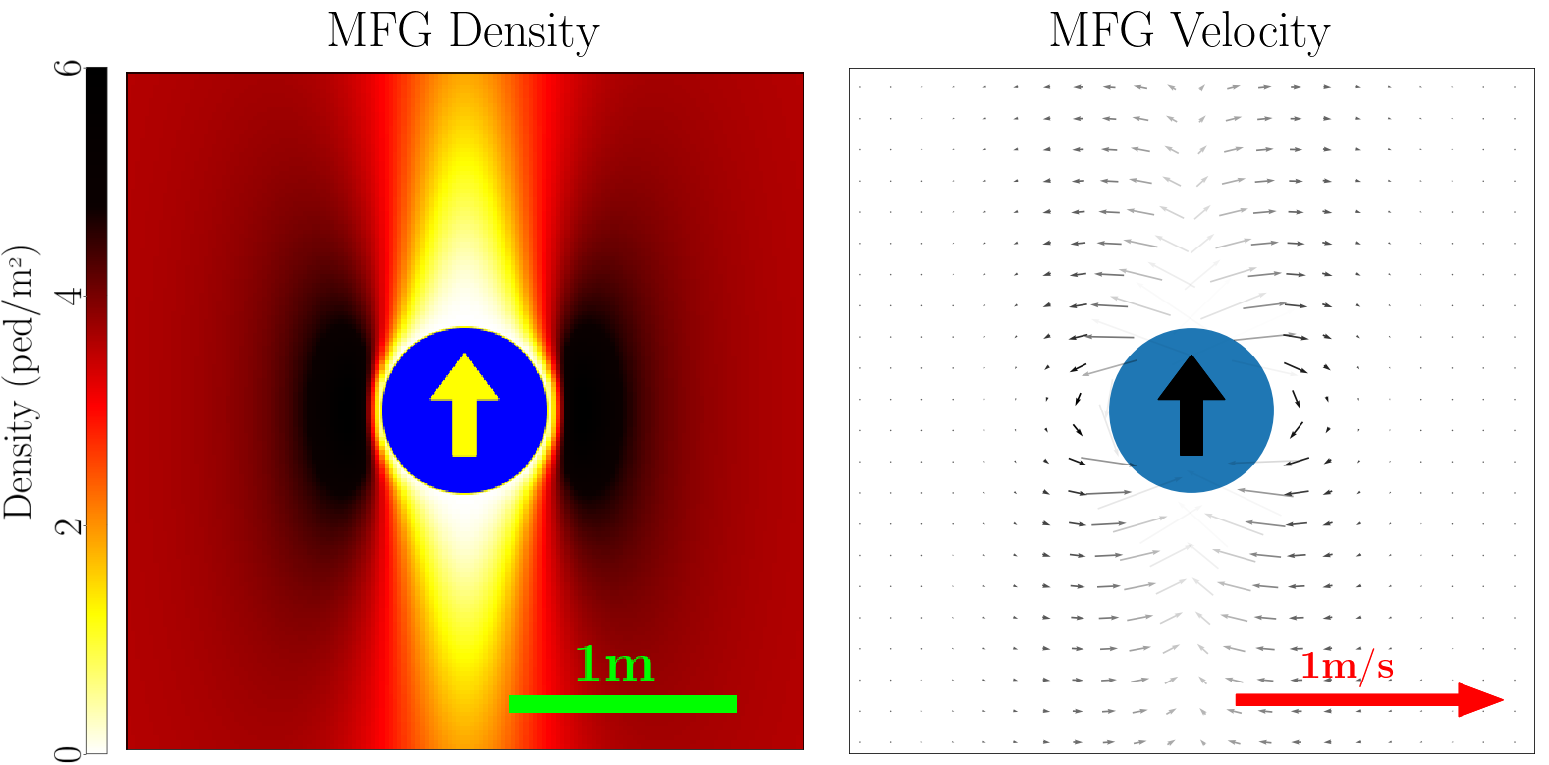}
    \caption{MFG simulation (\mb{for parameters $\xi = 0.15$ and $c_s = 0.11$, cf subsection \ref{subsec:understanding_mfg}}) of a static crowd being crossed by a cylindrical intruder. The density plot in the left panel presents the accumulation of agents on the cylinder sides and a depletion in front and behind the obstacle. The right panel depicts the velocity \tb{field}, where \tb{one can} see lateral displacement prior to the obstacle arrival and the return to lower density areas after its passage.}
    \label{fig:mfg_results}
\end{figure}

At this point we would like to show what the model is good for. Figure \ref{fig:mfg_results} \tb{displays} results \tb{of numerically solving the MFG system \eqref{eqn:MFG_sys} with parameters chosen to match the experimental data of Figure \ref{fig:exp_results}}.
\tb{This simulation exhibits} many of the main empirical features, starting from the accumulation on the sides and the depletion in front of and behind the obstacle. Moreover, the velocity field \tb{shows that the optimal MFG strategy involves the same lateral motion observed experimentally in \cite{Nicolas_Kuperman_mech_response}}. The model recovers remarkably well the overall experimental behaviour and the reason lies in its \tb{forward-backward structure}. \tb{In the MFG system \eqref{eqn:MFG_sys}, the backward HJB equation carries information both about the future positions of the intruder and about the future optimal configurations of the crowd. This informs the dynamics of the agents' density described by the coupled KFP equation, ensuring each individual in the crowd avoids the intruder with minimal discomfort, by minimising both the time spent in high density areas and their displacement.} Indeed, stepping asides means accepting to spend some time closer to others \tb{(up-front cost)}, but with the perspective of soon returning to a less crowded zone.

\subsection{Understanding MFG}  
\label{subsec:understanding_mfg}
\begin{figure}
    \centering
    \includegraphics[width =1\linewidth]{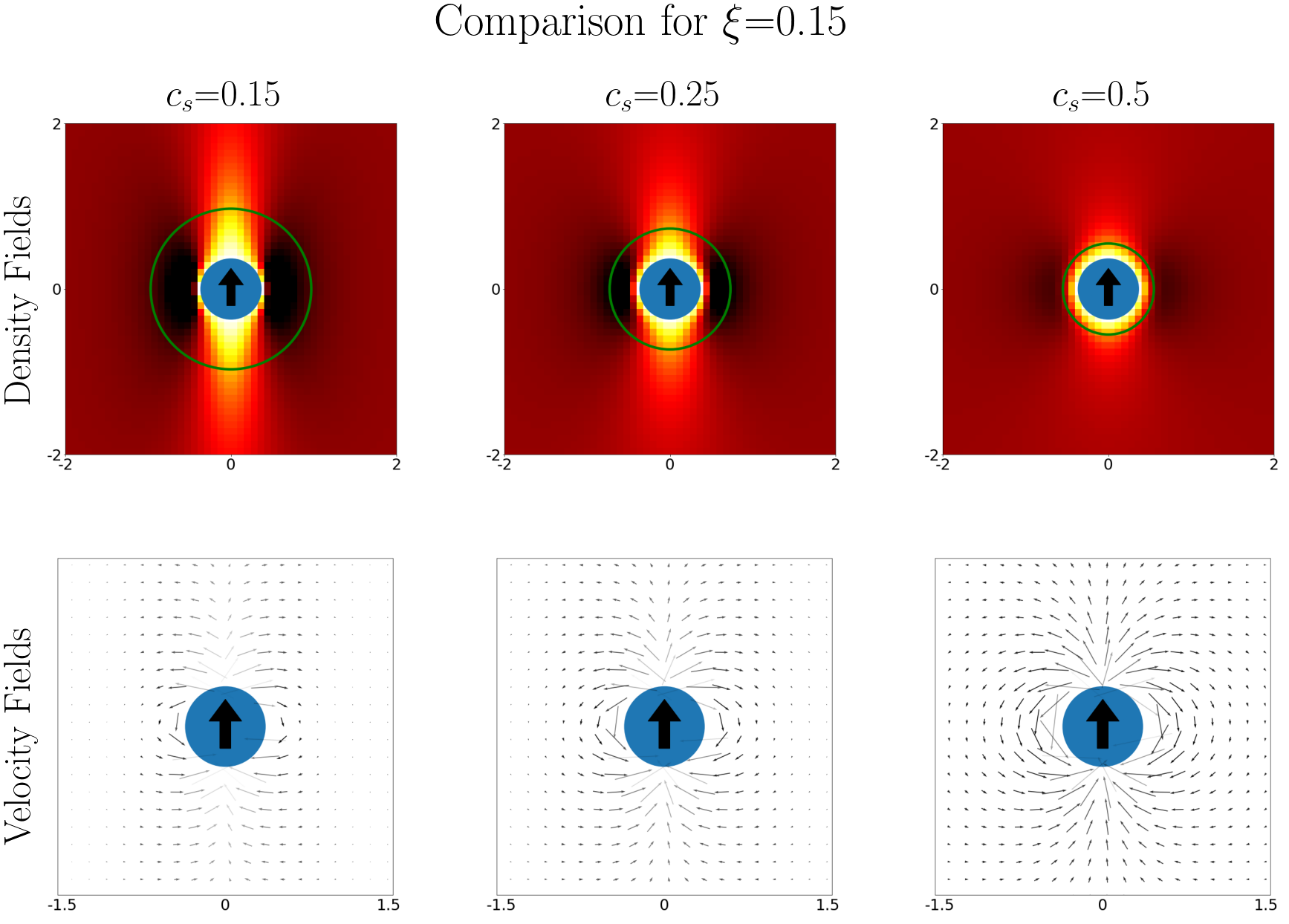}
    \caption{\tb{Density (top row) and velocity (bottom row) fields produced by solving the MFG equations with mean pedestrian density $m_0=2.5$, $R=0.37$, $v=0.6$, $\xi = 0.15$ and increasing values of $c_s = 0.15, 0.25, 0.5$. As $c_s$ increases, to avoid the intruder, agents abandon the transverse motion in favour of a circular one. Note how the green circle of diameter $l$ encompasses most of the perturbation caused by the intruder.}}
    \label{fig:scales}
\end{figure}

We \tb{should now clarify some additional features of the model. Most notably, by proper rescaling of Eqs \eqref{eqn:MFG_sys} ,} it is possible to reduce the number of parameters of the model to only two, namely the \emph{healing length} $\xi$ and the \emph{sound velocity} $c_s$ \cite{Bonnemain_Gobron_lax_connection,Bonnemain_Butano_ped_not_grains_players}
\begin{equation}
	\xi = \sqrt{\vline\frac{\mu\sigma^4}{2gm_0}\vline}, \qquad c_s = \sqrt{\vline\frac{gm_0}{2\mu}\vline}
\end{equation}
whose ratios with $R = 0.37m$, the cylinder radius, and $v =0.6m/s$\tb{, its velocity, respectively,} completely determine the phase space of the MFG solutions. 
Now, if we want to access radically different MFG strategies, we can do so by tuning $c_s$, whereas $\xi$ would only modulate how far from the cylinder pedestrians would be affected by its presence. That said, let us take a look at figure \ref{fig:scales}, ignoring the green circles  for the moment. The Figure shows three MFG simulations of the cylinder's passage, with the same $\xi$ as in Figure \ref{fig:mfg_results}\tb{, but} with three different values of $c_s$, ranging from left to right from $c_s/v \tb{\ll 1}$  to $c_s/v \sim 1$. \tb{Judging only by} the density \tb{fields} \tb{(first row)}, \tb{one merely} notices a difference in the extent of the perturbation, but the velocity fields \tb{(bottom row)} actually \tb{display} some \tb{noteworthy} change in  strategy. In fact, \tb{as $c_s/v$ decreases, moving (and especially moving fast) becomes more costly (in the game theoretical sense of Eq. \eqref{def:mean_cost}): agents prefer to anticipate the fast moving intruder and step away well in advance, even though} this means spending some time in a more congested area. On the other hand, for $c_s/v \sim 1$ agents do not mind moving \tb{and would rather remain} in low density areas, \tb{ultimately giving up their transverse motion in favour of} circulat\tb{ing} around the obstacle, as \tb{one may see} for $c_s = 0.5$. 

\tb{Besides outlining the optimal strategy, the two parameters $\xi$ and $c_s$ can be used to identify characteristic scales. The natural timescale obtained from their ratio $\tau = \xi/c_s$ represents the typical time the density needs to restore itself to its bulk value after a pointwise perturbation.} In our case however the intruder is not a point, and \mb{and it is not still. To take into account its radius and its velocity we propose} \mb{\sout{if we take into account its radius then we \tb{can define a new timescale $\tau_I = (\xi + R )/c_s $. Additionally, as the intruder is not still, but moving at speed $v$, }} $l = v\tau + R$ \mb{as \sout{is}} a good candidate to estimate the extent of the perturbation caused by the intruder. Indeed, looking again at Figure \ref{fig:scales}, the  green circles of radius $l$ give a scale of the perturbation caused by the intruder's motion. We should note that we have no general prescription to determine the perturbation scale }
for an arbitrary choice of   $U_0$. However, such scale should necessarily depend on the parameters $\xi$ and $c_s$.

\tb{Finally}, one of the common features of many agent-based models used to address \tb{the local navigation of crowds} is the need to choose each agent's anticipation time\tb{, resulting from a compromise between realism and numerical cost}. In these models the time \tb{over which each agent can anticipate} is chosen explicitly \tb{(or through some given rule)} and could be printed at each step of the \tb{simulation}. This is where MFG is radically different. \tb{The natural timescale of the game $T$ is not the anticipation time (and neither is $\tau$)}, \mb{but} 
the final time at which the \tb{game/}simulation ends\tb{;} it bears no consequences regarding the anticipatory dynamics  provided that it is large enough \tb{(cf the discussion on the stationary, or \emph{ergodic state} in \cite{Ullmo_Swiecicki_quadratic_mfg})}. Solving the MFG equations \tb{yields} the Nash equilibrium strategy of motion, describing how pedestrians would move optimally, \tb{as per} the cost functional \eqref{def:mean_cost}, for each point in space\tb{-time}\mb{\sout{. Therefore,}, meaning that, also} the time after which a pedestrian located in \tb{any given} point would start \tb{moving}, \tb{i.e.} the anticipation time, is  chosen optimally: it depends on the position and is not prescribed a priori. The anticipation dynamics is \mb{therefore} intrinsic to the MFG solution; we believe this is one of the key features of MFG, and one of the main reasons of its remarkable performance \tb{in this configuration}.
\section{Conclusion}
\label{sec:conc}
In this paper, we have presented a minimal model based on MFG, a mathematical \tb{framework describing the Nash equilibrium of an optimisation problem involving many interacting agents}. We have used such  model to simulate a dense crowd of pedestrians being crossed by a cylindrical intruder, \tb{\mb{and confronted the results} to experimental data}. Moreover, \tb{besides the model's ability to qualitatively capture  realistic behaviours,}  we have offered an overview of the main features  providing additional insight on why pedestrians behave the way they do in situations like the one studied here. We believe that the validity of our approach lies in the simplicity of its premises, i.e. that the agents' motion is derived only from the general individual preferences described by the cost functional \eqref{def:mean_cost}, without any prescription on the features it should display. The point is that the optimal motion is found as a consequence of such principles and not sought by means of ad hoc parameters. \tb{That said}, we do not believe \tb{pedestrians} solve a coupled system of non-linear PDEs while walking, but we support the claim that, at least in some simple situations, daily experience and evolution made it possible for the human brain to quickly find a collective solution to efficiently avoid obstacles while optimising the overall discomfort, a \tb{process} MFG can \tb{help} interpret and reproduce.

\bibliography{biblio}
\bibliographystyle{plain}

\end{document}